\documentclass[aps,prb,twocolumn,letterpaper,showpacs,preprintnumbers,amsmath,amssymb]{revtex4}

\usepackage{epsfig,amssymb}
\usepackage{graphicx}
\usepackage{dcolumn}
\usepackage{bm}
\usepackage{subfigure}
\usepackage{verbatim}

\begin{document}

\title{Structural phase transitions
in Ruddlesden-Popper phases of strontium titanate: \\
{\em ab initio} and
inhomogeneous Ginzburg-Landau approaches}

\author{Jeehye~Lee} \author{Tom\'{a}s~A.~Arias}
\affiliation{Laboratory of Atomic and Solid State Physics, Cornell
University, Ithaca, New York 14853}
\date{\today}

\begin{abstract} 
We present the first systematic {\em ab initio} study of
anti-ferrodistortive (AFD) order in Ruddlesden-Popper (RP) phases of
strontium titanate, Sr$_{1+n}$Ti$_n$O$_{3n+1}$, as a function of both
compressive epitaxial strain and phase number $n$.  We find all RP
phases to exhibit AFD order under a significant range of strains,
recovering the bulk AFD order as $\sim 1/n^2$.  A Ginzburg-Landau
Hamiltonian generalized to include inter-octahedral interactions
reproduces our {\em ab initio} results well, opening a pathway to
understanding other nanostructured perovskite systems.
\end{abstract}


\maketitle

Superlattices originating from various oxide perovskites are of great
interest due to their rich properties.  As examples, Sr$_2$RuO$_4$ of
the $n=1$ Ruddlesden-Popper (RP) family exhibits unconventional
superconductivity\cite{Sr2RuO4}, ferroelectricity in multicomponent
superlattices made of different perovskites (BaTiO$_3$,
SrTiO$_3$ and/or CaTiO$_3$) can be tuned by controlling the mixing
ratio \cite{Multicompo1, Multicompo2}, and, recently, even a
superconducting two-dimensional electron gas has been observed at the
interface of the SrTiO$_3$/LaAlO$_3$ superlattice\cite{STOLAO}.

The Ruddlesden-Popper series of structures,
$\mbox{A}_{n+1}\mbox{B}_n\mbox{O}_{3n+1}=(\mbox{A}\mbox{B}\mbox{O}_{3})_n(AO)$
(or, ``RP-$n$'') for $n=1, 2, \ldots$, is the simplest prototype of such
nano-ordered superlattices, consisting of a set of AO stacking fault
planes interspersed regularly between every $n$ layers of bulk
perovskite octahedra\cite{RuddlesdenPopper}.  
The RP-$n$ series most successfully synthesized
to date is that of strontium titanate, with phases up to $n=5$
reported to have been grown controllably with molecular beam
epitaxy(MBE)\cite{STORPExp}.  
The $n=\infty$ end member of the series,
bulk SrTiO$_3$, is also of great interest in its own right,
with highly tunable properties such as 
strain-switchable ferroelectricity\cite{FERoomTemp},
and even metallic and superconducting behavior under doping with
oxygen vacancies\cite{STO1, STO2}.

Despite its technological and scientific potential, the RP-$n$ series
of strontium titanate remains relatively unexplored.
The bulk material exhibits a
rich strain-temperature structural phase diagram including both
anti-ferrodistortive (AFD) and ferroelectric (FE) ordering and
various combinations thereof\cite{STOPhaseDiagram}, with non-FE AFD
order present at zero strain and temperature.  In contrast, previous
{\em ab initio} work has found there to be {\em no} AFD order at zero
strain and temperature in the RP-$n$ series for $n=1,2$, with the
expected bulk-like order only occurring in $n=3,4,5$ \cite{STORPCal4,
RP-AFD}.  However, the associated phase transitions with strain,
the underlying driving forces, the scaling of the behavior with $n$,
and the nature of the large-$n$ limit remains unexplored.  

This work addresses the above open issues by employing the {\em ab
initio} approach to study the AFD rotational transition as a function
of both distance between stacking faults $n$ and {\em epitaxial
strain} (biaxial strain applied in the plane of the SrO stacking
faults).  Such epitaxial strain is now accessible experimentally, with
substrates available to provide a substantial range of strains to tune
the material properties of strontium-titanate
structures\cite{STOPhaseDiagram, STORPstrain, STOstrain}.  For this
initial study, we here focus on compressive strains, which favor AFD
order with the rotation axis perpendicular to the stacking fault
planes.  This allows us to investigate the transition to bulk behavior
with increasing $n$, while freeing us from having to consider the more
complex configurations associated with the multiple, in-plane
rotational axes favored by tensile strains.

{\em Computational details --} All {\em ab initio} calculations below
employ the density-functional theory framework\cite{Payne} within the
local-density approximation (LDA) and represent the ionic cores with
norm-conserving Kleinman-Bylander pseudopotentials\cite{KBpot}.  A
plane-wave basis with a cutoff
energy of 30~H expands the Kohn-Sham orbitals, and Monkhorst-Pack
meshes\cite{kmesh} of $2\times2\times2$ or $2\times2\times1$ (for
RP-$n$ structures of $n\leq2$ or $n \geq 3$, respectively) sample
the Brillouin zone.  Finally, the resulting energy functional is
minimized by the analytically continued conjugate gradient
method\cite{Arias} within the {\tt DFT++} software\cite{Sohrab}.

We employ supercells composed of two bulk regions, containing $n$
strontium titanate bulk layers each, along with two extra SrO stacking fault
layers.  To properly represent the AFD order, each bulk layer contains two
structural units of strontium titanate.  
To represent epitaxial strain, we fix the in-plane lattice
parameter $a$ and relax the out-of-plane lattice constant $c$.
All calculations minimize the electronic wave functions to within
10~$\mu$H of the Born-Oppenheimer surface, the ionic coordinates until
all forces on individual atoms $\mid$F$\mid$ are less than 0.1~mH/B,
and the out-of-plane strain until it is determined to within
$\pm0.2\%$ (1~H$\approx$27.21~eV, 1~B$\approx$0.0529~nm).

\begin{figure}
\centering
\includegraphics[width=6.cm]{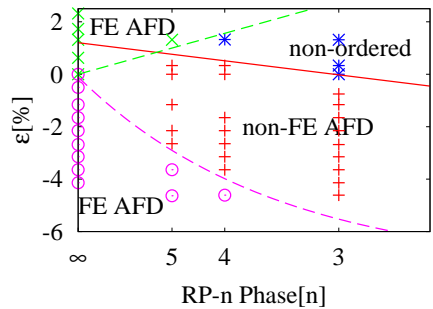}
\caption{(Color online) Structural phase diagram of
$\mbox{Sr}_{n+1}\mbox{Ti}_n\mbox{O}_{3n+1}$ as a function of
epitaxial strain and $n$: in-plane FE AFD order
($\times$), out-of-plane FE AFD order ($\circ$), out-of-plane non-FE AFD order (+), 
non-ordered (*), non-FE AFD to non-ordered phase boundary studied in
the text (solid line), additional phase boundaries (dashed lines).}
\label{RPPhaseDiagram}
\end{figure}

With the ultimate goal of comparing {\em ab initio} results to
experiments on epitaxially strained thin films, 
we define epitaxial strains as $\epsilon \equiv (a-a_B)/a_B$, 
where $a_B = 3.8376$~\AA~is the equilibrium lattice constant for cubic
(non-rotated) bulk SrTiO$_3$ within our computational framework.  We
find that small-$n$ phases under sufficiently low compressive strains
either exhibit AFD order with an out-of-plane rotation axis or remain
non-ordered (centrosymmetric).  To obtain reliable transition points
for low-$n$ phases and to extract information about bulk SrTiO$_3$
that can be connected directly to our results at low $n$, we sometimes
enter regions of phase space where the material exhibits other
instabilities.  Figure~\ref{RPPhaseDiagram} summarizes the phase space
we considered and the phases we found in each region.  
Under sufficiently high compressive or tensile strains,
AFD and FE orders coexist in the large-$n$ phases with the order
parameters oriented along either the out-of- or in- plane directions,
respectively, just as in bulk strontium titanate\cite{STOPhaseDiagram}.

Concerned here only with the transition between 
the out-of-plane non-FE AFD phase and
the non-ordered phase, for the calculations below, we suppress FE order
and in-plane AFD order by imposing both C$_2$ and mirror symmetry
about the out-of-plane direction.  Figure~\ref{RPPhaseDiagram} shows that, for
$n\leq 4$ and low strains, the ground-state phases are free of these
suppressed orders, so these symmetry restriction {\em in no way}
affect our conclusions for the $n\leq 4$ phases.  For the $n\geq 5$ phases,
the in-plane rotated order will likely become more thermodynamically
stable near the phase-transition point of interest to us.  Thus, for large
$n$, the rotated phases which we study may be only metastable near the
transition point.
Because we here are concerned with the nature of the non-FE AFD to
non-ordered phase boundary and how the spacing $n$ between stacking
faults affect this transition, the imposition of the above symmetry
restrictions again does not affect the central conclusions below.

{\em Sr$_2$TiO$_4$, $n=1$ phase --} As a first look at the AFD phase
transition under strain, we begin by noting that, by the symmetries of
the 00c$^-$ AFD ordering in the Glazer notation\cite{Glazer}, 
a single rotation angle $\theta$ serves to
define the state of rotation for each layer of octahedra.  In a very
simple picture in which the rotations in all layers in
each bulk region are the same (as is the case for the $n=1,2$ phases), we
would have a single order parameter and the usual Ginzburg-Landau form
for the free energy,
\begin{equation}
   F(\theta)=\frac{1}{2} A \theta^4 + \alpha (\epsilon - \epsilon_c) \theta^2 + F_0,
     \label{LandauGinzburg}
\end{equation}
where both $A$ and $\alpha$ are positive, and $\epsilon_c$ is the
critical strain of the transition, with the well-known solution 
that the ordered phase obtains for $\epsilon <
\epsilon_c$, with a nonzero rotation angle $\theta^2(\epsilon) =
\alpha(\epsilon_c - \epsilon)/A$ and an energy difference relative to
the centrosymmetric state ($\theta=0$) of $\sqrt{\Delta E(\epsilon)} =
\alpha(\epsilon_c - \epsilon)/\sqrt{2 A}$.  Thus, the {\em square} of
rotation angle and the {\em square-root} of the energy difference
should both vanish linearly as $\epsilon$ approaches $\epsilon_c$ from below.
(Here and below we regard compressive strains as $\epsilon<0$.)

   \begin{figure}
\centering
\includegraphics[width=6.cm]{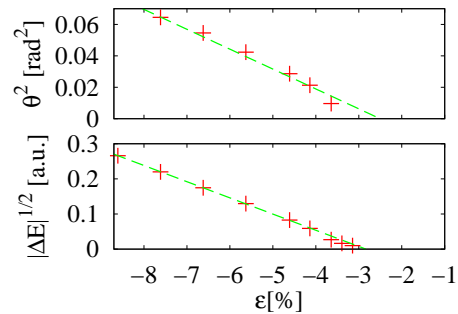}
\caption{(Color online) Square of rotational order parameter, $\theta^2$, versus epitaxial strain $\epsilon$ (upper panel); square-root of transition energy, $\sqrt{\Delta E}$, versus epitaxial strain $\epsilon$ (lower panel): {\em ab initio} results (crosses), linear fits (dashed lines).}
\label{RPN1data} 
\end{figure}

Figure~\ref{RPN1data} shows that our full {\em ab initio} results for
the $n=1$ RP phase exhibit quite clearly these classic signatures of a
second-order phase transition.  The extrapolated values for the
critical strain from the two observables, $\theta^2$ and $\sqrt{\Delta
E}$, are in surprisingly good agreement for such a simple model,
-2.2\% and -2.8\%, respectively. In either case, it is clear that the
critical strain is significantly below zero, consistent with the lack
of observation of AFD rotational order in the $n = 1$ phase near zero
strain in previous {\em ab initio} work\cite{STORPCal4}.  As we now
show, the critical strain passes through zero and becomes positive
with increasing $n$, eventually approaching a value associated with the bulk material.

\begin{figure}
     \centering
\subfigure{
    \includegraphics[width=5.6cm]{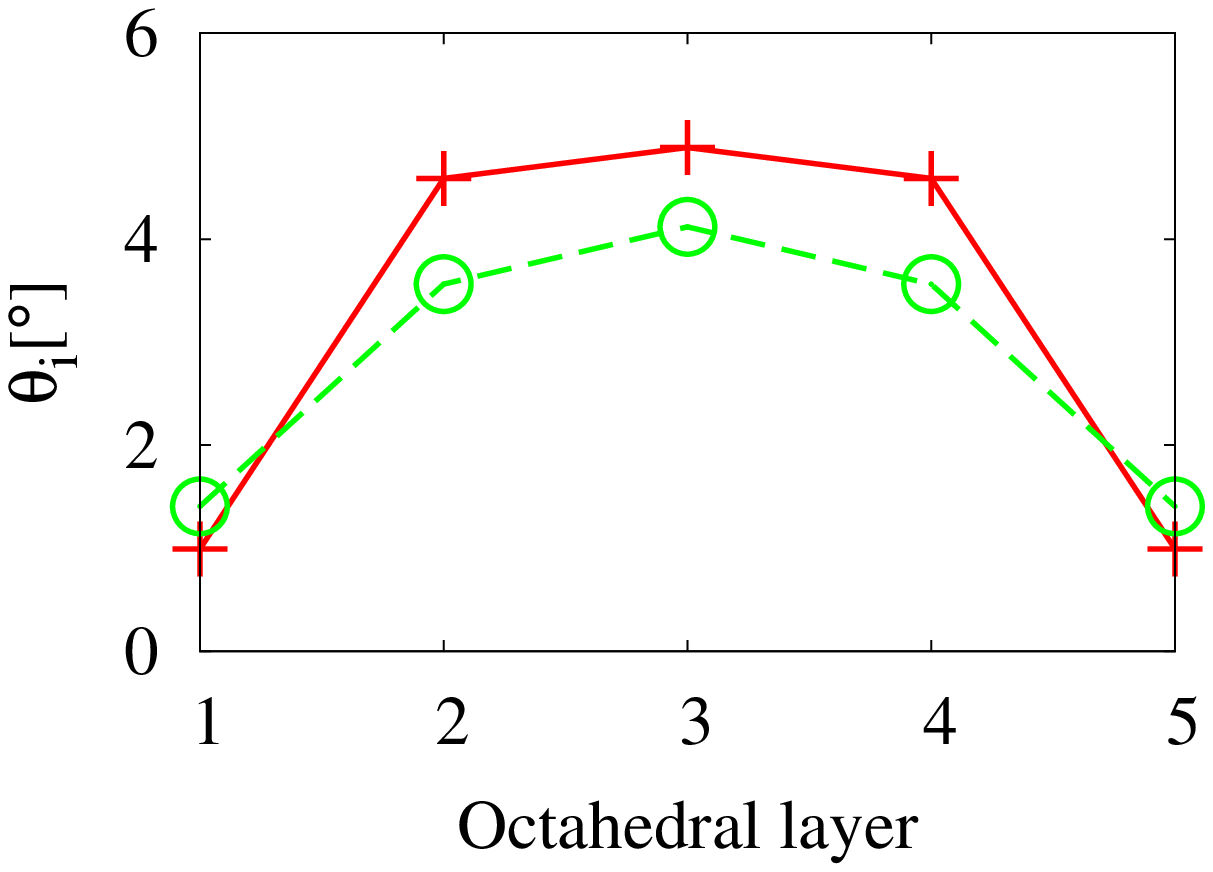}
    
}
\subfigure{
    \includegraphics[width=5.6cm]{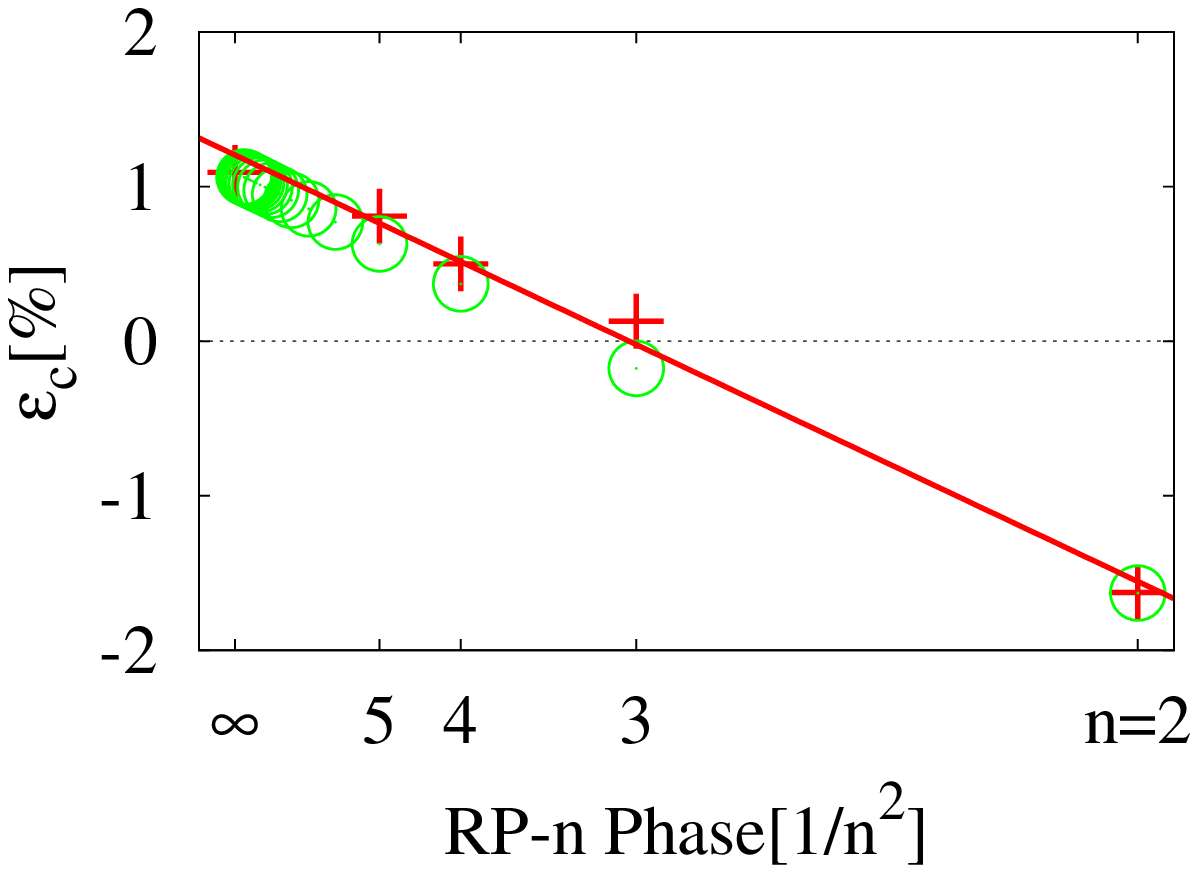}
}
\caption{(Color online) {\em Ab initio} (+) and Ginzburg-Landau model
  ($\circ$) results for layer-dependent rotation angles
  $\theta_i$ versus layer position for the $n=5$ phase at
  $\epsilon= 0.0\%$ (upper panel), and critical strain $\epsilon_c$
  versus {\em inverse-square} of RP phase number $n$ (lower panel).
  Lines in upper panel are guides to the eyes; solid line in lower panel is
  best fit to {\em ab initio} data.}
\label{fig:ModelDFTCompare}
\end{figure}

{\em Sr$_{n+1}$Ti$_n$O$_{3n+1}$, $n > 1$ phase --} For $n > 1$, multiple
perovskite layers exist between faults and different rotation angles
become possible for each layer (except for the $n=2$ phase 
where the two layers are identical by symmetry).  The upper panel of Figure
\ref{fig:ModelDFTCompare} displays the rotation angles in each
layer for the $n=5$ phase at zero strain.  These data exhibit the general
trend in all of our data, namely that the rotation angles are
smaller near the SrO stacking faults and approach the expected bulk
value toward the center of the bulk regions.  Associated with the
reduced rotation angle at the interface, we find an enhanced
stretching of the octahedra neighboring the stacking faults resulting in
rumpling of the SrO layers.  We also find off-center motion of the
titanium ions of these boundary octahedra toward the fault layers, so that each region of bulk
material, and thus the entire structure, exhibits no net ionic
electric dipole moment.  This suggests that the structure near the SrO
stacking faults is such that the boundary octahedra prefer stretching and
formation of a ionic electric dipole moment over rotation as a way to
relieve compressive stress, ultimately suppressing the rotational instability
near the faults.

To locate the AFD transition, for the order parameter
$\theta$, we take the rotation angle of the central one or two (if $n$ is odd
or even, respectively) perovskite layers, and extrapolate the linear
behavior of $\theta^2$ (which shows less numerical scatter than that
for $\sqrt{\Delta E}$), identifying the horizontal intercept as the
critical strain\cite{LDAerror}.  
The lower panel of Figure~\ref{fig:ModelDFTCompare} displays the
resulting critical strains for $n \geq 2$ as a
function of $1/n^2$.  With increasing $n$, the critical strain
steadily increases toward the bulk value as the RP-$n$ phases become
more like bulk.  The critical strains are {\em negative} for 
$n=1, 2$ and {\em positive} for $n\geq 3$, consistent with
and explaining the previous {\em ab initio} observations\cite{STORPCal4, RP-AFD} 
that only $n\geq 3$ phases exhibit AFD order 
when grown on a bulk SrTiO$_3$ substrate($\epsilon = 0$\%).
Finally, the clear linear behavior in the plot indicates that the
critical strain approaches the bulk value as $1/n^2$.

To explore the origin of the above behaviors, we generalize 
the above Ginzburg-Landau free-energy model to inhomogeneous
systems by including both inter-octahedral interactions and a
spatially-dependent tendency toward rotational instability,
\begin{equation}
  F(\{\theta_i\}) = \sum_{i}\left\{ \frac{1}{2}A\theta_i^4 + \alpha(\epsilon - \epsilon_i)\theta_i^2
   + \beta (\theta_i + \theta_{i-1})^2 \right\},
\label{Model}
\end{equation}
where $\theta_i$ is the octahedral rotation angle of the $i$-th layer,
$A$ and $\alpha$ are defined as above (Eq.~\ref{LandauGinzburg}),
$\epsilon_i$ represents an effective critical strain for the $i$-th
layer, and the third, inter-layer term imposes the
alternating sign ordering of $\theta_i$ associated with the Glazer
$00c^-$ order of the material.  To reflect the lessened tendency
towards rotation at the interfaces, $\epsilon_i$ should have more negative
values near the stacking faults, eventually approaching the critical
value for the bulk material $\epsilon_c(n=\infty)$ in the center of
the bulk regions.  In this initial work, we take $\epsilon_i$ to be
the bulk value $\epsilon_c(\infty)$ for all layers $i$ except those
neighboring the stacking faults, for which we will take a different
value, $\epsilon_i = \tilde{\epsilon}$, to be determined below.  The
standard stability analysis for this free energy is that there will be
a second-order phase transition when the Hessian, evaluated at
$\theta_i=0$ for all $i$, ceases to be positive definite.

To begin our analysis of the above free-energy form, we consider the
approach to the bulk behavior.  For sufficiently large $n$, one can
take the continuum limit, and the Hessian becomes 
$H=(\epsilon - \epsilon_c (\infty)) \theta(x) - (d^2 {\beta}/{\alpha}) {\partial^2
\theta}/{\partial x^2}$, where $\epsilon$ retains its meaning as the
applied strain, $d$ is the distance between adjacent layers and
$x$ is the position within the bulk region along the $c$-axis.  
Here, the layer-dependent rotation angle $(-1)^i \theta_i$ has now become the
smooth function $\theta (x)$.  
Finally, in passing to the continuum
limit, there is a boundary condition associated with the interfaces,
which for our simple model is
$(\epsilon_c(\infty) - \tilde{\epsilon})\theta(x) = (d
{\beta}/{\alpha}) {\partial \theta}/{\partial x}$, for $x=0,nd$.
The critical point where this continuum Hessian ceases to be positive
definite corresponds to the lowest eigenvalue of the standard
one-dimensional particle-in-a-box problem of quantum mechanics with a
modified boundary condition. The resulting critical strain is
$\epsilon_c = \epsilon_c(\infty) - (\beta q^2)/(\alpha n^2)$, where
$q$ satisfies the transcendental equation $\tan({q}/{2}) = \alpha\,
n\left({ \epsilon_c(\infty) - \tilde{\epsilon}}\right)/ (\beta q)$,
whose solution approaches a constant, $q \to \pi$, for large $n$.  
The continuum limit thus reproduces exactly the $1/n^2$ approach to
the bulk value observed in the {\em ab initio} data.  
The slope $\gamma$ observed in the {\em ab initio} data can now be identified as
a combination of parameters from the Ginzburg-Landau
Hamiltonian, namely $\gamma = (\beta/\alpha)\pi^2$.

Using the observed slope in the {\em ab initio} data as well as data
from the bulk and $n=2$ phases, we now determine completely the
parameters in the Ginzburg-Landau model.  In the bulk phase, the last
term of Eq.~\ref{Model} vanishes due to the AFD order, and the
analysis becomes identical to that for $n=1$ phase, in which $A$,
$\alpha$ and $\epsilon_c(\infty)$ can all be determined from the
linear fits to the $\theta^2$ and $\sqrt{\Delta E}$ versus $\epsilon$
data.  When $n = 2$, by symmetry, the Hamiltonian again assumes a form
of Eq.~\ref{LandauGinzburg}, but with $\tilde{\epsilon}$ instead of
$\epsilon_c(\infty)$, so that we directly read off
$\tilde{\epsilon} = \epsilon_c(n=2)$.  The final parameter remaining
is $\beta$, which we determine from the above result for $\gamma$ in
terms of $\beta$ and $\alpha$, taking $\gamma$ from the slope of the
best-fit line to the {\em ab initio} data (solid line, lower panel of
Figure~\ref{fig:ModelDFTCompare}).

The open circles in the lower panel of
Figure~\ref{fig:ModelDFTCompare} represent the critical strains from
eigenvalues of the above Hessian for inhomogeneous systems, with
parameters extracted from the {\em ab initio} results as above.  The
clear linear behavior confirms our analytic prediction that the
critical strain in the model approaches the bulk value as $1/n^2$.
The upper panel of Figure~\ref{fig:ModelDFTCompare} presents an even
more compelling case for the model free energy.  It compares {\em ab
initio} and model-Hamiltonian results for a quantity not included in
the parameter fitting at all, the layer-dependent rotation angles
$\theta_i$ for the $n=5$ RP-$n$ phase at $\epsilon=0$\% strain. The
agreement is impressive considering the simplicity of the model
and the fact that the model is not fit to these
quantities.

\begin{figure}
 \centering
\includegraphics[width=6cm]{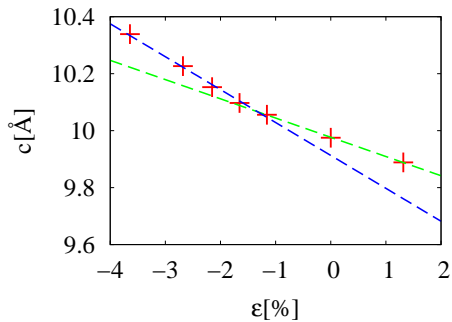}
\caption{(Color online) Out-of-plane lattice constant versus the biaxial strain $\epsilon$:
{\em ab initio} results (+), linear fits to
data on either side of the transition (dashed lines).
}
\label{fig:CvsepsB_N2}
\end{figure}
{\em Experimental signatures ---} In terms of an experimentally
observable signature of this transition, we find that, associated with
the rotational ordering, there is a significant expansion of the
out-of-plane lattice constant $c$, by an amount which should be
readily detectable with x-ray techniques.  Figure~\ref{fig:CvsepsB_N2}
shows the lattice constant $c$ of the $n=2$ phase as a function of
$\epsilon$ through the AFD transition point. The transition is clearly
evident where the slope of the curve changes substantially between
$\epsilon=-1\%$ and $-2\%$ strain. The lattice constant $c$ for the
AFD phase is significantly larger ($>0.6\%$ for strains $\epsilon <
-2.6$\%, about 1\% strain beyond the transition) than the extrapolated
value from the non-rotated phase. A perhaps more practical approach is
to grow various members of the RP-$n$ series on substrates of {\em
fixed} epitaxial strain $\epsilon$ and to observe the dependence
$c(n)$ of the out-of-plane lattice constant as a function of $n$,
looking again for a kink at the transition point.  
We stress that the numerical values given here apply specifically to zero temperature.

{\em Conclusion --- } This work presents a detailed {\em ab initio}
study of the effects of strain on the rotational instability in the
Ruddlesden-Popper series in strontium titanate for $n=1,\cdots, 5$.
We find a similar second-order structural phase transition to what is
seen in bulk strontium titanate, but with a critical point displaced
by $\sim 1/n^2$ from the bulk value.  The key microscopic mechanism
for the $n$-dependence is the interplay of the local
AFD coupling between neighboring octahedra layers
with a depression of the rotational instability for octahedra
immediately neighboring the SrO stacking faults.
It appears that the depression is associated with the extra freedom which the neighboring
octahedra have to distort by formation of local ionic electric dipole
moments and by rumpling of the extra SrO layers.
A simple Ginzburg-Landau free-energy expression generalized for
inhomogeneous systems by inclusion of nearest-neighbor
inter-octahedral interactions captures the essential features of the
{\em ab initio} results for the $n$-dependence of the transition.
Despite its simplicity, this free-energy expression gives reasonable
predictions even for quantities to which it is not fit, such as the
distribution of rotation angles in the bulk regions of the RP phase.
This success suggests that inclusion of inter-octahedral interactions
into more general bulk Hamiltonians, particularly those with
ferroelectric degrees of freedom, will be a fruitful direction to
pursue in future studies of nano-structured and superlattice
perovskite materials.

This work was supported by the Cornell Center for Materials
Research (CCMR) under the NSF MRSEC program (DMR-0520404).
The authors are grateful to Craig Fennie for many fruitful discussions.
\bibliography{paper3}

\begin{thebibliography}{21}
\bibitem[{LDA()}]{LDAerror}
\bibinfo{note}{The LDA overbinding error may lead to an uncertainty in
  transition points as the AFD rotational instability is known to be
  overestimated.}
\end{thebibliography}
\end{document}